# The Optimised Theta Method

José Augusto Fioruci$^a$, Tiago Ribeiro Pellegrini$^b$, Francisco Louzada$^c$ and Fotios Petropoulos$^d$

$^a$ Department of Statistics, Federal University of São Carlos, Brazil

$^b$ Department of Mathematics and Statistics, University of New Brunswick, Canada

$^c$ Department of Applied Mathematics and Statistics, University of São Paulo, Brazil

$^d$ Lancaster Centre for Forecasting, Lancaster University, UK

**Abstract**

Accurate and robust forecasting methods for univariate time series are very important when the objective is to produce estimates for a large number of time series. In this context, the Theta method called researchers attention due its performance in the largest up-to-date forecasting competition, the M3-Competition. Theta method proposes the decomposition of the deseasonalised data into two "theta lines". The first theta line removes completely the curvatures of the data, thus being a good estimator of the long-term trend component. The second theta line doubles the curvatures of the series, as to better approximate the short-term behaviour. In this paper, we propose a generalisation of the Theta method by optimising the selection of the second theta line, based on various validation schemes where the out-of-sample accuracy of the candidate variants is measured. The recomposition process of the original time series builds on the asymmetry of the decomposed theta lines. An empirical investigation through the M3-Competition data set shows improvements on the forecasting accuracy of the proposed optimised Theta method.

## 1 Introduction

The development of accurate, robust and reliable forecasting methods for univariate time series is very important when a large number of time series are involved in the modelling and forecasting process. In many industrial settings it is very common to work with a large line of products; thus, efficient sales and operational planning (S&OP) heavily depends on accurate forecasting methods.

Despite the advantages on automatic model selection algorithms (Hyndman *et al.*, 2002; Hyndman & Khandakar, 2008; Poler & Mula, 2011; Assimakopoulos, 1995), there is still the need for accurate extrapolation methods that will form the pool of methods from which one can select the most appropriate for the data in hand. To that end, the impact of empirical forecasting competitions is very important. The M3-Competition (Makridakis & Hibon, 2000) is the largest up-to-date forecasting competition, including more than 3,000 time series (referring to various types of data and different frequencies) and 24 participants (methods, selection approaches, and specialised software). A relatively simple approach, the Theta method, proposed by Assimakopoulos & Nikolopoulos (2000) (hereafter A&N), managed to outperform all other participants.

The core idea behind the A&N is the fully exploitation of the data, through a decomposition which takes place on the seasonally adjusted data. This decomposition aims to the magnification of the short- and the long-term movements of the data. The movements are captured via the so-called "theta lines". The Theta method, as applied by A&N in the M3-Competition, decomposes the data into exactly two theta lines with pre-fixed parameters, directly reflecting to the specific degree of enhancement of the short and long-term behaviour of the data. Later on Hyndman & Billah (2003) proposed an approximation (Nikolopoulos *et al.*, 2011) of the Theta method, referring to this as Simple Exponential Smoothing with drift (SES-d). The performance of the Theta method has been confirmed by other empirical studies (for example: Nikolopoulos *et al.*, 2012; Petropoulos & Nikolopoulos, 2013). Moreover, Hyndman & Billah (2003) and Thomakos & Nikolopoulos (2014) provided theoretical insights of the Theta method. Despite these advances, we believe that the Theta method deserves more attention from the forecasting community, given its simplicity and superior forecasting performance.

One key aspect of the Theta is that, by definition, this method is dynamic. One can choose different theta lines and combine the produced forecasts with equal or unequal weights. However, A&N limit this important property, by fixing the theta coefficients to have predefined values. Thus, the Theta method, as implemented in the M3-Competition, is limited in the sense that it focuses only on specific information of the data. On the contrary, if the selection of the appropriate theta lines was carried out through optimisation, then the method could focus on the information that is actually important.

In this work we extend the A&N method by optimally selecting the theta line that best describes the short-term movements of the series, maintaining the long-term component. In order to select a theta line for the short-term behaviour, a loss function based on prediction errors over a validation sample is minimised for each series. The combination of the forecasts derived from the two theta lines is performed using appropriate weights, which ensure the recomposition of the original time series. An empirical study using the M3-Competition database is conducted in order to decide amongst various validation schemes. The results reveal improvements with regards to the forecasting accuracy of the proposed optimised Theta method, outperforming several benchmarks as well as the original implementation by A&N.

The paper is organised as follows. Section 2 describes the original Theta method of A&N. Section 3 presents the Optimised Theta method and its estimation process. Section 4 presents the forecasting performance of the proposed method, compared to a list of widely used benchmarks. The evaluation includes more than 3,000 time series. Section 5 presents our final comments and plans for future research.

## 2 The Theta method

The intuitive idea of the Theta method Assimakopoulos & Nikolopoulos (2000) has its basis on the decomposition of the original time series based on their local curvatures. The decomposition process is made through a theta coefficient (denoted by the greek letter $\theta$), $\theta \in \mathbb{R}$, which is applied on the second differences of the data. When $\theta < 1$, the second differences are reduced resulting in the better approximation of the long-term behaviour of the series Assimakopoulos (1995). If $\theta$ is equal to 0, the decomposed line is transformed into a straight line with constant slope, which is the simple linear regression on time. On the other hand, when $\theta > 1$ the local curvatures are increased, magnifying the short-term movements of the time series (Assimakopoulos & Nikolopoulos, 2000). The decomposed lines produced are called theta lines, denoted here by $Z_t(\theta)$. These lines have the same mean value and slope with the original data. However, the local curvatures are filtered out or enhanced, depending on the value of the $\theta$ coefficient.

In other words, the decomposition process has the advantage of exploiting more information in the data that usually cannot be completely captured and modelled though the extrapolation of the original time series. The theta lines can be regarded as new time series and are forecasted separately, using an appropriate forecasting method. Once the extrapolation of each theta line has been completed, recomposition takes place through a combination scheme as to calculate the point forecasts of the original time series. Combining has long been considered as a useful practice in the forecasting literature (for example: Makridakis & Winkler, 1983; Clemen, 1989; Sayed *et al.*, 2009; Martins & Werner, 2012) and, so, its application in the Theta method is expected to result to more accurate and robust estimates.

Nikolopoulos *et al.* (2012) provided, among several empirical results, a very intuitive and simple formula to calculate the theta lines as a linear combination of the data and a fitted time trend given by,

$$Z_t(\theta) = \theta y_t + (1-\theta)(\hat{\alpha} + \hat{\beta} t), \qquad (1)$$

where $y_t$ is the original time series at time $t = 1, ..., n$, and $\hat{\alpha}$ and $\hat{\beta}$ are the usual least squares estimators. Under this point of view, the theta lines can be interpreted as a function of the linear regression model applied directly to the data.

The steps to build the Theta method of A&N are as follows (Assimakopoulos & Nikolopoulos, 2000):

Step 0. *Seasonality Test:* The time series is tested for statistically significant seasonal behaviour, using an auto-correlation function.

Step 1. *Deseasonalization:* If needed, the time series is deseasonalized via the classical decomposition method, assuming a multiplicative relationship of the seasonal component.

Step 2. *Decomposition:* The time series is decomposed into two theta lines, the linear regression line $Z_t(0)$ and the theta line for $Z_t(2)$.

Step 3. *Extrapolation:* The linear regression line is extrapolated in the usual way, while the second line is extrapolated via the simple exponential smoothing (SES) method.

Step 4. *Combination:* The forecasts produced from the extrapolation of the two lines are combined with equal weights (50%-50%).

Step 5. *Reseasonalization:* The forecasts are reseasonalized, if they were deseasonalized in Step 1.

This approach, based on two theta lines with ad-hoc values for the $\theta$ coefficients and equal weight for the recomposition of the final forecasts, resulted in the best performance for the largest up-to-date forecasting competition, the M3 competition (Makridakis & Hibon, 2000).

Generalisations of the Theta method have been considered over the time. For example, Nikolopoulos & Assimakopoulos (2005) and Petropoulos & Nikolopoulos (2013) argue for the use of more theta lines, $\theta \in \{-1, 0, 1, 2, 3\}$, as to extract even more information from the data. Empirical evidence suggests that the consideration of more/different theta lines can result in improvements compared to the original Theta method. However, a formalised procedure on selecting appropriate theta lines is yet to be proposed.

Moreover, Constantinidou *et al.* (2012) and Petropoulos & Nikolopoulos (2013) suggested the use of unequal weights in the recomposition procedure of the final forecasts. This is an intuitively appealing approach, as asymmetric weights, which are directly linked with the forecast horizon, are likely to offer a better approximation of the short- and long-term components. However, there is a strong justification to use equal weights in Step 4 where the theta lines $Z_t(0)$ and $Z_t(2)$ are combined. The justification is given by the following equation

$$
\begin{aligned}
0.5 Z_t(0) + 0.5 Z_t(2) &= 0.5(\hat{\alpha} + \hat{\beta}t) + 0.5[2y_t - (\hat{\alpha} + \hat{\beta}t)] \\
&= y_t,
\end{aligned}
\quad (2)
$$

So, by definition, the decomposition of the original series in $Z_t(0)$ and $Z_t(2)$ suggests the use of equal weights, if the original signal is to be reconstructed. In other words, the use of weights that derived directly from the decomposition procedure (the corresponding $\theta$ coefficients) will provide a better starting point.

Building on these findings, we propose an optimised version of the Theta method that generalises the decomposition proposed by A&N. We propose a formalised procedure to select appropriately the $\theta$ coefficient of the second theta line. An appropriate computational methodology is considered to keep high forecasting performance with low computational effort in order to provide accurate forecasts for a large number of time series.

## 3  The Optimized Theta method

For an observed time series $y_1, \ldots, y_n$ a generalisation of the theta lines combination by Theta method can be written by the following equation,

$$
Y_t = \omega Z_t(\theta_1) + (1 - \omega) Z_t(\theta_2) \quad (3)
$$

for $t = 1, ..., n$, where $\omega \in [0, 1]$ is the weight parameter. Note that, if $\theta_1 = 0, \theta_2 = 2$ and $\omega = 0.5$ the method (3) results in A&N method. But, note that equation (3) does not satisfy the equalities $Y_t = y_t$, $t = 1, \ldots, n$ for any values of $\theta_1, \theta_2$ and $\omega$. However, Theorem 1 ensures that this can be achieved if we consider the weight parameter $\omega$ as a function of the parameters $\theta_1$ and $\theta_2$ if, and only if, $\theta_1 \leq 1$ and $\theta_2 \geq 1$. The resulting weight is given by

$$\omega := \omega(\theta_1, \theta_2) = \frac{\theta_2 - 1}{\theta_2 - \theta_1}, \tag{4}$$

defining $\omega(1, 1) = 1$. The Theorem 1 ensures that this solution is unique as well.

Note that, if $\theta_1 = 0$ and $\theta_2 = 2$, equation (4) results in $\omega = 0.5$. Therefore, equations (3) and (4) allow to construct a generalisation of the Theta method that maintain the propriety of recomposition of the original time series for any theta lines $Z_t(\theta_1)$ and $Z_t(\theta_2)$, with $\theta_1 \leq 1 \leq \theta_2$.

**Theorem 1** *The linear system given by $Y_t = y_t$ for all $t = 1, \ldots, n$, where $Y_t$ is given by the equation (3), has a single solution if, and only if, the weight $\omega$ is given by the equation (4), with the restrictions $\theta_1 \leq 1$ and $\theta_2 \geq 1$.*
Proof: *Firstly, note that equation (1) can be rewritten as*

$$Z_t(\theta) = \theta(y_t - \hat{\alpha} - \hat{\beta}t) + \hat{\alpha} + \hat{\beta}t.$$

*So the linear system given by*

$$Y_t = y_t, \quad \text{for all } t \in \{1, \ldots, n\}$$

*implies*

$$\begin{aligned}
0 &= Y_t - y_t \\
&= \omega[\theta_1(y_t - \hat{\alpha} - \hat{\beta}t) + \hat{\alpha} + \hat{\beta}t] + (1 - \omega)[\theta_2(y_t - \hat{\alpha} - \hat{\beta}t) + \hat{\alpha} + \hat{\beta}t] - y_t \\
&= \omega\theta_1(y_t - \hat{\alpha} - \hat{\beta}t) + (1 - \omega)\theta_2(y_t - \hat{\alpha} - \hat{\beta}t) - (y_t - \hat{\alpha} - \hat{\beta}t) \\
&= [\omega\theta_1 + (1 - \omega)\theta_2 - 1](y_t - \hat{\alpha} - \hat{\beta}t), \quad \text{for all } t \in \{1, \ldots, n\}.
\end{aligned}$$

*Note that, $y_t - \hat{\alpha} - \hat{\beta}t$ is the residue of the linear regression model and cannot be zero for all $t \in \{1, \ldots, n\}$. Hence, the upper equations are true if, and only if, $\omega\theta_1 + (1 - \omega)\theta_2 - 1 = 0$, which implies in $\omega = (\theta_2 - 1)/(\theta_2 - \theta_1)$, as we want to show.*

□

In order to retain the long-term component of the data, in this work we focus in the case $\theta_1 = 0$ and $\theta_2 = \theta$, where $\theta \geq 1$ is the only parameter which will be optimised based on a given loss function. The Figure (1) present the weights as a function of $\theta$. Thus, the decomposition for the Optimised Theta method (OTM) is given by

$$Y_t = \left(1 - \frac{1}{\theta}\right)(\hat{\alpha} + \hat{\beta}t) + \frac{1}{\theta}Z_t(\theta),$$

with $\theta \geq 1$ and the forecasts for $k$ steps ahead of $t$ are given by

$$\hat{Y}_{t+k|t} = \left(1 - \frac{1}{\theta}\right)\left[\hat{\alpha} + \hat{\beta}(t+k)\right] + \frac{1}{\theta}\hat{Z}_{t+k|t}(\theta), \tag{5}$$

where $\hat{Z}_{t+h|t}(\theta)$ is the extrapolated theta line via a defined forecasting method, where one possibility is to use SES as in A&N. However, one could move from SES by selecting an appropriate extrapolation method (Petropoulos *et al.*, 2014) based on the features of the $Z_t(\theta)$.

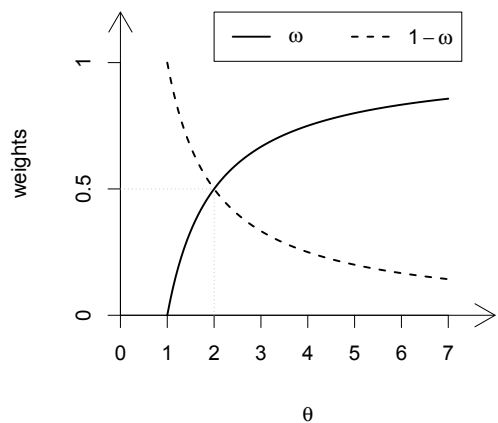

Figure 1: The relationship between the weights and $\theta$ parameter of the OTM.

An important propriety of the OTM is that when $\theta = 1$, which implies that $Z_t(1) = y_t$, the forecasting vector given by equation (5) will be equal to $\hat{Y}_{t+k|t} = \hat{Z}_{t+h|t}(1)$, which in this case it is simply the forecast of a defined extrapolated method directly applied on $y_t$. This means that the OTM extends the extrapolation method of the second theta line. For instance, if the SES method is used to produce forecasts for $Z_t(\theta)$, then when $\theta = 1$ the OTM falls to the SES method. When $\theta > 1$, then OTM acts as an extension of SES, by adding a long-term component.

### 3.1 Estimation

An intuitive idea to estimate the best value of $\theta$ for forecasting over the time series $y_1, \ldots, y_n$ is to minimise a loss function of predictions errors in the time series. One possibility is minimising a loss function of the prediction errors with one-step-ahead for each time $t$ as follows:

$$l_1(\theta) = \sum_{t=1}^{n-1} g(\hat{Y}_{t+1|t}, y_{t+1}), \tag{6}$$

where $\hat{Y}_{t+k|t}$ is given by equation (5) and $g(.,.)$ is a symmetric positive function, such as the Square Error function, defined as $SE(a,b) = (a-b)^2$, the Absolute Error function, defined as $AE(a,b) = |a-b|$ or the symmetric Average Percentage Error function, defined as $sAPE(a,b) = 2|a-b|/(|a|+|b|)$.

The OTM can be estimated by minimising the loss function (6) using some numerical optimisation algorithm, such as the L-BFGS-B algorithm (Byrd *et al.*, 1995). However, focusing on the minimisation of the one-step-ahead predictions can be regarded as sub-optimal, when the purpose is forecasting for multiple steps ahead (Fildes & Petropoulos, 2014). Moreover, this process may become computationally expensive, when forecasts are required for a large number of time series.

An intuitive idea to estimate the value of $\theta$ that yields better forecasts is to divide the time series in two parts, where the first part is used to fit the method (training sample) and the second part is used to evaluate the accuracy of the method (validation sample). This process is known as fixed origin evaluation (Tashman, 2000). To fix the notation, let $n$ be the sample length, let $n_1$ denote the length of first block (or origin) and $h$ be the number of required forecasts (forecasting horizon). Figure 2 shows an example of this process. In this example, the time series in black is the in-sample with length $n = 64$ and the objective is to forecast the eight values in red ($h = 8$). The idea is to split the in-sample in two blocks, fit the OTM in the first block containing $n_1 = 44$ values (training sample) and choose the value of the $\theta$ coefficient that gives the best forecasts in block 2 with length 20 (validation sample). In essence, one has to test different values of $\theta$ and produce forecasts for the validation sample with each one of them. Then the prediction errors in the validation sample are calculated and the method with the minimum average error (optimal $\theta$ coefficient) is selected. The selected method is fitted for the full sample (training and validation) and the required out-of-sample forecasts are produced. The differences of the actuals (red values) and the out-of-sample forecasts refer to the actual forecast errors of our method.

The process of fixed origin evaluation focuses on a single origin, which is a good option in order to reduce the computational effort; however, this process is very susceptible to be influenced by local characteristics. A possibility to overcome this problem is to update the origin recursively. For this purpose, a rolling origin evaluation scheme (Tashman, 2000) is considered. Let $n_i$ denote the origin of predictions (last position of the fitted period) in step $i = 1, \ldots, p$, where $p \geq 1$ is the number of origin updates. Let $m \geq 1$ be the number of movements ahead of the origin in each step and $H \geq 1$ be the number of predictions (test period) in each step. Obviously, we have that $1 < n_1 < n_2 < \cdots < n_p < n$, with $n_{i+1} = n_i + m$, for $i = 1, \ldots, p-1$. Moreover, it is not difficult to see that the maximum number of updates is

$$p_{\max} = 1 + \left\| \frac{n - n_1}{m} \right\|, \tag{7}$$

where $\|x\|$ denotes the largest integer less than $x \in \mathbb{R}$. So the resulting loss function is

$$l(\theta) = \sum_{i=1}^{p} \sum_{j=1}^{min(H, n-ni)} g(\, y_{n_i+j} \,,\, \hat{Y}_{n_i+j|n_i} \,), \tag{8}$$

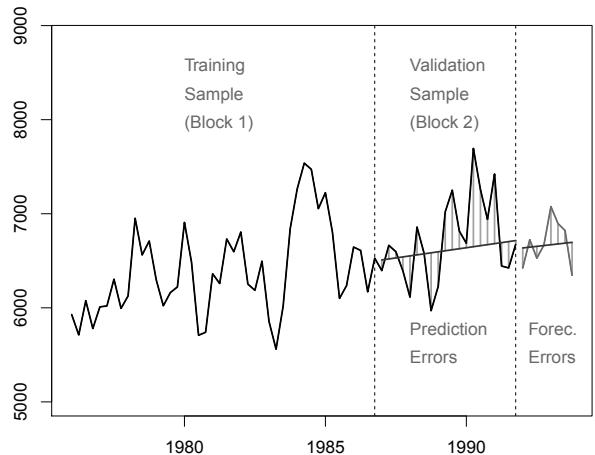

Figure 2: Example of fixed origin evaluation.

where $1 \leq p \leq p_{\max}$.

From this point forward we will refer to equation (8) as Generalised Rolling Origin Evaluation (GROE). Note that, the only parameters necessary to compute the GROE for some forecasting methods are $p, m, H$ and $n_1$. Figure 3 presents a graphical illustration of this generalised evaluation procedure. Both rolling origin and fixed origin evaluation are special cases of the GROE. Equation (8) corresponds to rolling origin evaluation if $m = 1$ and $H \geq n - n_1$, while a fixed origin evaluation can be achieved by setting $m = H \geq n - n_1$. For $n_1 = 2$ and $m = H = 1$ the equation (8) corresponds to equation (6).

When $m = H$ the number of movements from the origin and the number of predictions is the same in each update, leading to a structure of non-overlapping blocks. This structure enables to cover a fairly long test period with few forecasting updates. An example of this process is presented in Figure 4 for GROE with $p = 3$ origins, where the first origin is $n_1 = 25$ and the numbers of advances and predictions are 13 for each step. Therefore, the others origins are $n_2 = 38$ and $n_3 = 51$. Note that, the evaluation process occurs with only 3 fittings of the OTM, since the predictions in block 2 are calculated by fitting the method in the block 1; the predictions in block 3 are derived using blocks 1 and 2; finally the predictions of block 4 are produced using the blocks 1, 2 and 3. We expect that the process presented in Figure 4 is less susceptible to local characteristics than the process presented in Figure 2.

For fixed values of $n_1, m$ and $H$ the $\theta$ estimation can be directly obtained by brute-force search in the loss function (8), which is fast and robust in our case. Other search algorithms may be considered. However, considering the L-BFGS-B algorithm we faced slowness, which can be a disadvantage when one is working with a large number of time series. In order to be able to use brute-force search, the parametric space of $\theta$ needs to be limited. This makes sense for this method, since for large values of $\theta$ the corresponding theta lines will have a high degree of variability, and, thus, a non-predictable behaviour. Another important aspect to take into account is that the theta line does not present great variation for a small neighbourhood of theta, hence in practice the parametric space can be considered discrete and limited. For instance, by limiting $\theta \in [1, 5]$ the estimation is done choosing the $\theta \in \Theta = \{1, 1.5, 2, 2.5, 3, 3.5, 4, 4.5, 5\}$ that gives the lowest prediction error over the validation sample(s) according to the loss function (8).

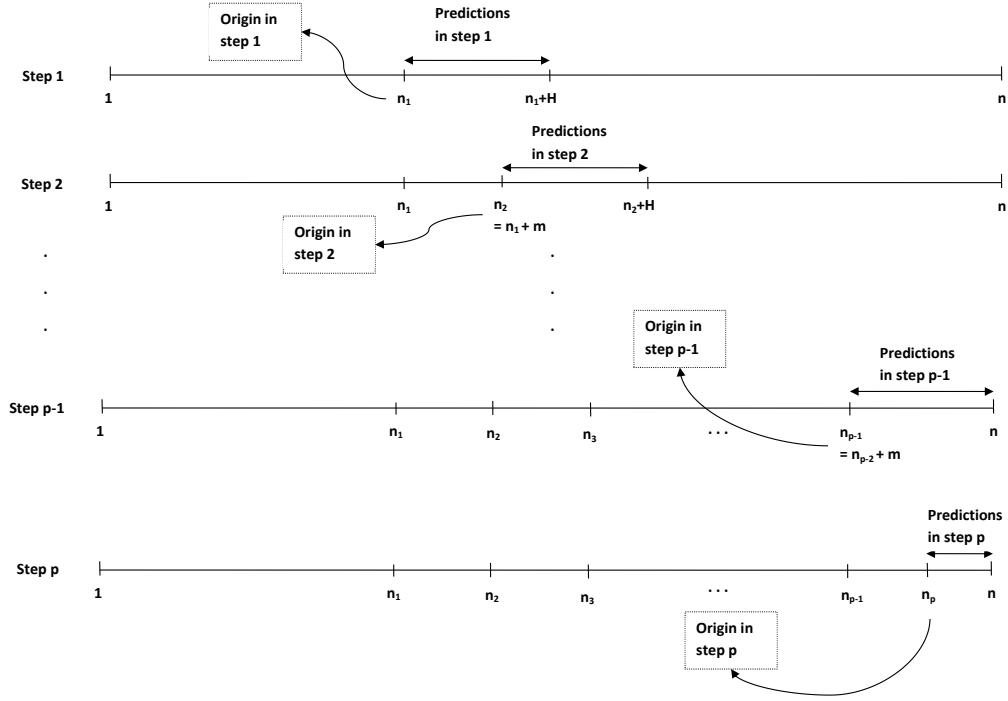

Figure 3: A graphical illustration of the Generalised Rolling Origin Evaluation.

## 3.2 Algorithm for the Optimised Theta Model

For fixed values of $n_1, m, H, h$, and $\Theta$ the steps to build the OTM are as follows:

Step 0. *Seasonality Testing:* The time series is tested for statistically significant seasonal behaviour.

Step 1. *Deseasonalization:* If needed, the time series is deseasonalized via the classical decomposition method, assuming a multiplicative relationship of the seasonal component.

Step 2. *Estimation:* Let $\hat{\theta}$ be the value of $\Theta$ that gives the lowest value of the loss function (8).

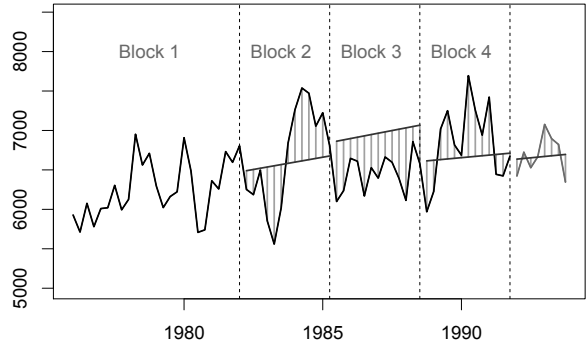

Figure 4: Example of Generalised Rolling-Origin with $n_1 = 25$ and $m = H = 13$.

Step 3. *Decomposition:* The time series is decomposed into two theta lines, the linear regression line $Z_t(0)$ and the theta line for $Z_t(\hat{\theta})$.

Step 4. *Extrapolation:* The linear regression line is extrapolated in the usual way, while the second line is extrapolated via SES.

Step 5. *Combination:* The forecasts produced from the extrapolation of the two lines were combined using equation (5).

Step 6. *Reseasonalization:* The forecasts are reseasonalized, if they were deseasonalized in Step 1.

# 4 Empirical Study and Application

In order to obtain empirical evidence about the best choice for the GROE parameters ($p, m, H$ and $n_1$) and the cost function $g$, in this section we present a empirical study for eight alternatives of how to fix these parameters. More details about the database are presented in the Table 1.

Table 1: M3-Competition data per frequency and requested forecast horizon.

| Frequency | Forecasting Horizon (h) | Number of time series |
|:---:|:---:|:---:|
| Yearly | 6 | 645 |
| Quarterly | 8 | 756 |
| Monthly | 18 | 1428 |
| Other | 8 | 174 |
| Total | | 3003 |

The hypotheses about the GROE parameters were made taking into account only the size ($n$) and forecasting horizon ($h$) of each time series. We consider the sizes $h$ and $2h$ for the prediction window, which implies two levels for the first origin parameter, i.e., $n_1 = n - h$ and $n_1 = n - 2h$. For the number of movements of the origin in each step we consider the levels $h$, $\|h/2\|$, $\|h/3\|$ and 1. In order avoid slowness during the estimation process, the maximum number of origin updates in $h$ are limited, i.e, it is considered $p = \min(p_{\max}, h)$ for all hypotheses. The parameter of the number of predictions in each step ($H$) is fixed as $h$. The combinations of all parameters levels are enumerated from (a) to (h) in Table 2. For some time series (192 yearly and 52 quarterly) the level $n - 2h$ implies $n_1 < 4$. In this cases we set $n_1 = 4$.

Table 2: Sets of parameters for the different GROE approaches.

| Approaches | $p$ | $m$ | $H$ | $n_1$ |
|---|---|---|---|---|
| (a) | 1 | $h$ | $h$ | $n - h$ |
| (b) | 2 | $\|h/2\|$ | $h$ | $n - h$ |
| (c) | 3 | $\|h/3\|$ | $h$ | $n - h$ |
| (d) | $h$ | 1 | $h$ | $n - h$ |
| (e) | 2 | $h$ | $h$ | $n - 2h$ |
| (f) | 4 | $\|h/2\|$ | $h$ | $n - 2h$ |
| (g) | 6 | $\|h/3\|$ | $h$ | $n - 2h$ |
| (h) | $h$ | 1 | $h$ | $n - 2h$ |

Note that case (a) is equal to the fixed origin evaluation, cases (d) and (h) refer rolling origin evaluations alternatives and case (e) refers to non-overlapping blocks. The others approaches, cases (b),(c),(f) and (g), were chosen to be in-between special cases of GROE. Case (h) is the only one where $p < p_{\max}$ for some time series.

For the cost function $g$, three metrics are considered: SE, AE and sAPE. To evaluate the out-of-sample performance we use two established metrics of accuracy, the symmetric Mean Absolute Percentage Error (sMAPE) and the Mean Absolute Scared Error (MASE). The sMAPE metric is defined as the mean of sAPE, i.e.,

$$sMAPE = \frac{200}{h} \sum_{i=1}^{h} \frac{|y_{n+i} - \hat{Y}_{n+i|n}|}{|y_{n+i}| + |\hat{Y}_{n+i|n}|},$$

and the MASE metric, proposed by Hyndman & Koehler (2006), is the mean of AE divided by the mean of the first difference in the time series, i.e.,

$$MASE = \frac{n-1}{h} \frac{\sum_{i=1}^{h} |y_{n+i} - \hat{Y}_{n+i|n}|}{\sum_{t=2}^{n} |y_t - y_{t-1}|}.$$

To compare the performance of the proposed OTM, we choose the classical Theta method of A&N, which corresponds to the OTM with $\theta = 2$ and other widely used benchmark methods, for which further details are given in Table 3. Amongst them, we choose methods of the exponential smoothing family (ETS), as well as the automatic algorithms implemented in the *forecast* package by Hyndman & Khandakar (2008).

Table 3: The benchmark methods used in the currect study.

| Method | Reference | Description |
|---|---|---|
| Naive |  | Random Walk |
| Naive2 | Makridakis & Hibon (2000) | Seasonal Random Walk |
| SES | Brown (1956) | ETS(A,N,N) |
| Holt or | Holt (1957) | ETS(A,A,N) for non seasonal data or |
| Holt-Winters | Winters (1960) | ETS(A,A,M) for seasonal data |
| Damped or | Gardner & McKenzie (1985) | ETS(A,Ad,N) for non seasonal data or |
| Seasonal Damped |  | ETS(A,Ad,M) for seasonal data |
| ETS | Hyndman & Khandakar (2008) | ETS automatic algorithm |
| ARIMA | Hyndman & Khandakar (2008) | Automatic Arima based on AICc |

To produce forecasts for the $Z_t(\theta)$ we considered three methods: SES, Holt and Damped. We will present in detail the results obtained by using the SES estimator and we will shortly discuss on the results derived using Holt and Damped methods for extrapolating the second theta line.

The empirical part of this study was implemented using the open-source statistical software provided by R Core Team (2013) and the packages *forecast, Mcomp* and *parallel*. The computer used for this task was equipped with a processor Intel Xeon E7-2870, 32GB of RAM which was operating on Linux Debian 3.2.46.

The empirical results are summarised in Tables 4 and 5 for the metrics sMAPE and MASE, respectively. The best results in each set are marked in bold and the cases where OTM obtained superior results from all benchmarking methods are highlighted with gray cells.

The replication of the Theta method implemented in this research was, as expected, the method that obtained overall the best performance across the benchmarks. Any numerical differences with the published results in the M3-Competition (Makridakis & Hibon, 2000) are due to the use of different pre-fixed theta coefficients and extrapolation methods for each frequency of the data (Nikolopoulos *et al.*, 2011). The small differences in the monthly data are deriving from the use of different software and optimisation functions. The performance of the Theta method is followed by that of Damped, ETS and ARIMA. Examining the results with regards to the various frequencies, the only case were the Theta method is not the best performer is the segment of "other" data. In this case the Damped method and the ETS algorithm obtained the best results for sMAPE and MASE respectively.

With regards to the performance of the OTM, the three cost functions, sAPE, AE and SE, obtained very similar results, regardless the accuracy metric for out-of-sample evaluation (sMAPE or MASE). When all data are considered, variation in the results was less than or equal to 0.09 for the sMAPE metric and less than or equal to 0.01 for the MASE metric. Comparing across the different cases for the sets of GROE's parameters, approaches (a) to (d) were generally better than the approaches (e) to (h), while the approach (d) and (h) provided the best and the worst results respectively.

Table 4: Empirical results on the performance of benchmark methods and OTM. Out-of-sample metric: sMAPE.

| Methods | Yearly | Quarterly | Monthly | Other | All | Time (min) |
|---|---|---|---|---|---|---|
| Benchmarking methods | | | | | | |
| Theta | **16.73** | **9.30** | **13.88** | 4.92 | **13.09** | 0.20 |
| Naive | 17.88 | 11.32 | 18.18 | 6.30 | 16.58 | 0.08 |
| Naive 2 | 17.88 | 11.07 | 17.23 | 6.30 | 15.88 | 0.07 |
| SES | 17.77 | 10.90 | 16.23 | 6.28 | 15.15 | 0.05 |
| Holt/Holt-Winters | 19.55 | 11.29 | 15.97 | 4.59 | 15.15 | 0.50 |
| Damped | 17.15 | 10.33 | 14.22 | **4.26** | 13.52 | 0.89 |
| ETS | 17.93 | 9.86 | 14.29 | 4.28 | 13.57 | 13.10 |
| ARIMA | 17.58 | 9.98 | 15.35 | 4.54 | 14.30 | 5.74 |

| Estimation Approach | Yearly | Quarterly | Monthly | Other | All | Time (min) |
|---|---|---|---|---|---|---|
| OTM with sAPE as cost function for the validation | | | | | | |
| (a) | 16.42 | 9.21 | 13.78 | 4.63 | 12.96 | 1.44 |
| (b) | 16.22 | 9.18 | 13.73 | 4.63 | 12.90 | 2.66 |
| (c) | 16.28 | **9.14** | 13.70 | 4.65 | 12.88 | 4.26 |
| (d) | 16.21 | **9.14** | **13.66** | 4.66 | **12.85** | 15.99 |
| (e) | 16.08 | 9.28 | 13.80 | **4.56** | 12.95 | 2.63 |
| (f) | **15.96** | 9.37 | 13.76 | 4.58 | 12.93 | 5.01 |
| (g) | 15.99 | 9.41 | 13.77 | 4.62 | 12.95 | 8.05 |
| (h) | 16.11 | 9.41 | 13.93 | 4.65 | 13.07 | 15.22 |

| Estimation Approach | Yearly | Quarterly | Monthly | Other | All | Time (min) |
|---|---|---|---|---|---|---|
| OTM with AE as cost function for the validation | | | | | | |
| (a) | 16.41 | 9.23 | 13.81 | 4.63 | 12.99 | 1.30 |
| (b) | 16.32 | 9.19 | 13.75 | 4.63 | 12.93 | 2.39 |
| (c) | 16.37 | **9.16** | 13.73 | 4.65 | 12.92 | 3.72 |
| (d) | 16.33 | **9.16** | **13.65** | 4.66 | **12.86** | 13.93 |
| (e) | 16.16 | 9.32 | 13.80 | **4.54** | 12.97 | 2.27 |
| (f) | **16.02** | 9.38 | 13.82 | 4.56 | 12.98 | 4.35 |
| (g) | 16.13 | 9.43 | 13.82 | 4.60 | 13.00 | 7.00 |
| (h) | 16.16 | 9.44 | 13.92 | 4.64 | 13.07 | 12.77 |

| Estimation Approach | Yearly | Quarterly | Monthly | Other | All | Time (min) |
|---|---|---|---|---|---|---|
| OTM with SE as cost function for the validation | | | | | | |
| (a) | 16.41 | 9.29 | 13.84 | 4.62 | 13.02 | 1.20 |
| (b) | 16.37 | 9.25 | 13.77 | 4.62 | 12.96 | 2.17 |
| (c) | 16.34 | 9.23 | 13.67 | 4.65 | 12.89 | 3.43 |
| (d) | 16.31 | **9.21** | **13.63** | 4.62 | **12.85** | 12.91 |
| (e) | 16.20 | 9.34 | 13.89 | **4.55** | 13.04 | 2.10 |
| (f) | 16.20 | 9.41 | 13.83 | 4.56 | 13.01 | 4.05 |
| (g) | 16.20 | 9.45 | 13.83 | 4.60 | 13.02 | 6.49 |
| (h) | **16.19** | 9.44 | 13.90 | 4.63 | 13.06 | 12.23 |

Table 5: Empirical results on the performance of benchmark methods and OTM. Out-of-sample metric: MASE.

| | Benchmarking methods | | | | | |
|---|---|---|---|---|---|---|
| | Yearly | Quartely | Monthly | Other | All | Time (min) |
| Theta | **2.77** | **2.08** | **2.12** | 2.27 | **2.19** | 0.20 |
| Naive | 3.17 | 2.39 | 2.60 | 3.09 | 2.64 | 0.08 |
| Naive 2 | 3.17 | 2.76 | 3.30 | 3.09 | 3.19 | 0.07 |
| SES | 3.17 | 2.36 | 2.51 | 3.09 | 2.58 | 0.05 |
| Holt/Holt-Winters | 3.14 | 2.45 | 2.93 | 1.91 | 2.84 | 0.50 |
| Damped | 2.95 | 2.32 | 2.71 | 1.78 | 2.64 | 0.89 |
| ETS | 3.21 | 2.41 | **2.12** | **1.77** | 2.27 | 13.10 |
| ARIMA | 2.99 | 2.24 | 2.15 | 1.87 | 2.24 | 5.74 |

| | OTM with sAPE as cost function for the validation | | | | | |
|---|---|---|---|---|---|---|
| Estimation Approach | Yearly | Quartely | Monthly | Other | All | Time (min) |
| (a) | 2.68 | 2.06 | 2.06 | 2.03 | 2.13 | 1.44 |
| (b) | 2.66 | 2.04 | 2.05 | 2.03 | 2.11 | 2.66 |
| (c) | 2.67 | **2.02** | 2.04 | 2.04 | 2.11 | 4.26 |
| (d) | 2.65 | **2.02** | **2.02** | 2.04 | **2.09** | 15.99 |
| (e) | 2.66 | 2.10 | 2.07 | **1.98** | 2.13 | 2.63 |
| (f) | **2.64** | 2.12 | 2.11 | 1.99 | 2.16 | 5.01 |
| (g) | **2.64** | 2.13 | 2.12 | 2.01 | 2.17 | 8.05 |
| (h) | 2.65 | 2.13 | 2.16 | 2.03 | 2.20 | 15.22 |

| | OTM with AE as cost function for the validation | | | | | |
|---|---|---|---|---|---|---|
| Estimation Approach | Yearly | Quartely | Monthly | Other | All | Time (min) |
| (a) | 2.68 | 2.06 | 2.06 | 2.03 | 2.13 | 1.30 |
| (b) | 2.66 | 2.04 | 2.05 | 2.03 | 2.11 | 2.39 |
| (c) | 2.66 | 2.03 | 2.04 | 2.04 | 2.11 | 3.72 |
| (d) | 2.66 | **2.02** | **2.02** | 2.04 | **2.09** | 13.93 |
| (e) | 2.66 | 2.10 | 2.07 | **1.97** | 2.13 | 2.27 |
| (f) | **2.63** | 2.12 | 2.11 | 1.98 | 2.16 | 4.35 |
| (g) | 2.64 | 2.13 | 2.13 | 2.00 | 2.18 | 7.00 |
| (h) | 2.66 | 2.14 | 2.16 | 2.02 | 2.21 | 12.77 |

| | OTM with SE as cost function for the validation | | | | | |
|---|---|---|---|---|---|---|
| Estimation Approach | Yearly | Quartely | Monthly | Other | All | Time (min) |
| (a) | 2.67 | 2.07 | 2.06 | 2.02 | 2.13 | 1.20 |
| (b) | 2.66 | 2.05 | 2.05 | 2.02 | 2.11 | 2.17 |
| (c) | 2.66 | **2.03** | 2.04 | 2.04 | 2.10 | 3.43 |
| (d) | 2.65 | **2.03** | **2.02** | 2.02 | **2.09** | 12.91 |
| (e) | 2.65 | 2.11 | 2.09 | **1.98** | 2.14 | 2.10 |
| (f) | 2.64 | 2.12 | 2.12 | 2.00 | 2.17 | 4.05 |
| (g) | **2.63** | 2.13 | 2.12 | 2.00 | 2.17 | 6.49 |
| (h) | 2.65 | 2.13 | 2.15 | 2.02 | 2.20 | 12.23 |

If we restrict the analysis only in the level $n - h$ for the GROE parameter $n_1$ (i.e., if we consider only the estimation approaches (a) to (d)), then the results of the proposed method (OTM) are better than all benchmarking methods for "yearly", "quarterly" and "monthly" time series frequencies, as well as for "all" time series frequencies. The other level for the GROE parameter $n_1$ ($n - 2h$) seems to be more suitable only for time series with "other" frequencies.

Overall, the results presented in the Table 4 reveal that the OTM provided better results than the classic Theta method. Focusing on the results obtained using the sAPE as cost function and the sMAPE as the out-of-sample error metric, we also consider the Multiple Comparisons with the Best test (MCB) for "all" time series frequencies in order to compare statistically the OTM with the Theta method. We consider the eight estimation approaches of the OTM and the Theta method, in total nine methods, which are compared with each other. In this test, for each method a rank interval is constructed (see Koning *et al.* (2005) for details). When the rank intervals of a par of methods are not overlapping, the null hypothesis of same accuracy is rejected in favor the alternative hypothesis of different accuracy.

The average ranks and the rank intervals of each method are presented in the Figure 5, which also present a comparison of the average rank of each method with the best average rank, by adopting a significance level equals to 5%. In contrast to what had been pointed out by the accuracy metrics sMAPE and MASE, the approaches (b) and (c) provided better rank results than (d), while the OTM with approach (c) gave the minimum average rank. However, the OTM with approaches (a) to (d) are not significantly different from the best. According to the average rank, the worst estimation approach is (h), a result which is in line with the standard out-of-sample accuracy evaluation provided by sMAPE and MASE. Most importantly, when the rank interval of approach (h) is compared with the rank intervals of any other approaches, the intervals are not overlapping, meaning that the approach (h) is statistically worse than any other estimation approach. However, the OTM with approach (h) is still statistically better than the classic Theta method. This result corroborate the evidence on the OTM accuracy performance.

In terms of the computational times achieved, the OTM is, as expected, more computational intensive than the original Theta method. However, the calculation times of the OTM are comparable to that of the two automatic model selection algorithms (`ets()` and `auto.arima()`) implemented in the *forecast* package (Hyndman & Khandakar, 2008).

As mentioned, in addition to the SES estimator for extrapolating the $Z_t(\theta)$, we reproduced the empirical study using Holt and Damped as forecasting methods. The results for the Holt method are always worse than SES, while Damped showed in general improvements for time series with "other" frequency; however, the computational effort increased significantly.

# 5 Concluding remarks

In this paper we proposed a generalisation of the Theta method, namely OTM. The OTM is the first generalisation proposed in the literature that provides accurate reconstruction of the original time series. To estimate the optimal Theta method, we proposed the minimisation of a loss function based on different validation schemes. The M3-Competition data were used to test the alternative estimation approaches. The performance of the OTM was compared against the original implementation of the Theta method, as well as against widely used forecasting benchmarks.

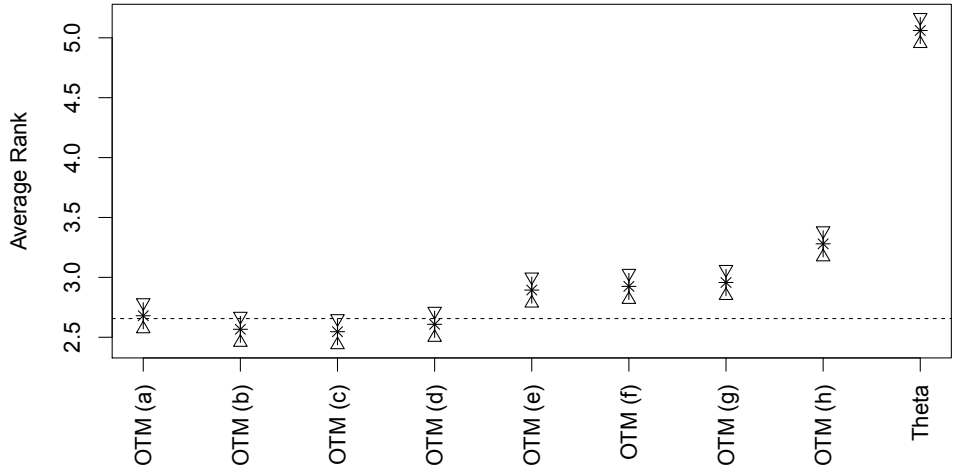

Figure 5: MCB intervals from OTM approaches and the original Theta method.

The proposed OTM demonstrated improvements in the forecasting performance of the classic Theta method, and it proved to be the most accurate and robust extrapolation approach across all benchmarks. Moreover, all variants of OTM were significantly better than the original Theta method, according to the average ranking of the different approaches.

We believe that this study has very important implications for decision makers. We show that the new optimised version of the Theta method improves over the original approach, significantly when ranks are taken into account. Keeping in mind that the original Theta method was already a very good and robust estimator for fast-moving demand, the new OTM achieves even higher levels of forecasting accuracy, which can be directly translated into profits. So, the OTM is able to provide better baseline statistical estimates, which can then be combined with judgmental overrides (Franses & Legerstee, 2011) to produce the final (operational) forecasts.

Future work on the OTM will be based on selecting different theta lines directly linked with the forecasting horizon, making distinct choices for the short- and the long-term. Moreover, future research should focus on the appropriate selection of extrapolation methods for the theta lines, rather than considering pre-fixed estimators, such as linear regression line for $Z_t(0)$ and SES for $Z_t(\theta)$.

# References

Assimakopoulos, V. (1995). A sucessive filtering technique for identifying long-term trends. *Journal of Forecasting*, **14**, 35–43.

Assimakopoulos, V. & Nikolopoulos, K. (2000). The theta model: a decomposition approach to forecasting. *International Journal of Forecasting*, **16**(4), 521 – 530.

Brown, R. G. (1956). *Exponential Smoothing for Predicting Demand*. Cambridge, Massachusetts: Arthur D. Little Inc.

Byrd, R. H., Lu, P., Nocedal, J. & Zhu, C. (1995). A limited memory algorithm for bound constrained optimization. *SIAM Journal on Scientific Computing*, **16**, 1190–1208.

Clemen, R. T. (1989). Combining forecasts: A review and annotated bibliography. *International Journal of Forecasting*, **5**, 559–583.

Constantinidou, C., Nikolopoulos, K., Bougioukos, N., Tsiafa, E., Petropoulos, F. & Assimakopoulos, V. (2012). A neural network approach for the theta model. *Information Engineering, Lecture Notes in Information Technology*, **25**, 116–120.

Fildes, R. & Petropoulos, F. (2014). An evaluation of simple versus complex selection rules for forecasting many time series. *Journal of Business Research*, **forthcoming**.

Franses, P. H. & Legerstee, R. (2011). Combining sku-level sales forecasts from models and experts. *Expert Systems with Applications*, **38**, 2365–2370.

Gardner, J. E. S. & McKenzie, E. (1985). Forecasting trends in time series. *Management Science*, **31**, 1237–1246.

Holt, C. C. (1957). *Forecasting seasonals and trends by exponentially weighted averages*. O. N. R. Memorandum 52/1957. Pittsburgh: Carnegie Institute of Technology.

Hyndman, R. J. & Billah, B. (2003). Unmasking the theta method. *International Journal of Forecasting*, **19**, 287–290.

Hyndman, R. J. & Khandakar, Y. (2008). Automatic time series forecasting: the forecast package for R. *Journal of Statistical Software*, **27**, 1–22.

Hyndman, R. J. & Koehler, A. B. (2006). Another look at measures of forecast accuracy. *International Journal of Forecasting*, **22**, 679–688.

Hyndman, R. J., Koehler, A. B., Snyder, R. D. & Grose, S. (2002). A state space framework for automatic forecasting using exponential smoothing methods. *International Journal of Forecasting*, **18**, 439–454.

Koning, A. J., Franses, P. H., Hibon, M. & Stekler, H. O. (2005). The M3 competition: Statistical tests of the results. *International Journal of Forecasting*, **21**(3), 397–409.

Makridakis, S. & Hibon, M. (2000). The M3-competition: results, conclusions and implications. *International Journal of Forecasting*, **16**, 451–476.

Makridakis, S. & Winkler, R. L. (1983). Averages of forecasts: Some empirical results. *Management Science*, **29**, 987–996.

Martins, V. L. M. & Werner, L. (2012). Forecast combination in industrial series: A comparison between individual forecasts and its combinations with and without correlated errors. *Expert Systems with Applications*, **39**, 11479–11486.

Nikolopoulos, K. & Assimakopoulos, V. (2005). Fathoming the theta model. In *25th International Symposium on Forecasting, ISF, San Antonio, Texas, USA*. unknown.

Nikolopoulos, K., Assimakopoulos, V., Bougioukos, N., Litsa, A. & Petropoulos, F. (2011). The Theta model: An essential forecasting tool for supply chain planning. *Advances in Automation and Robotics, Lecture Notes in Electrical Engineering*, **123**, 431–437.

Nikolopoulos, K., Thomakos, D., Petropoulos, F., Litsa, A. & Assimakopoulos, V. (2012). Forecasting S&P 500 with the Theta model. *International Journal of Financial Economics and Econometrics*, **4**, 73–78.

Petropoulos, F. & Nikolopoulos, K. (2013). Optimizing Theta model for monthly data. In *Proceedings of the 5th International Conference on Agents and ArtificialIntelligence*.

Petropoulos, F., Makridakis, S., Assimakopoulos, V. & Nikolopoulos, K. (2014). 'Horses for Courses' in Demand Forecasting. *European Journal of Operational Research*, pages –.

Poler, R. & Mula, J. (2011). Forecasting model selection through out-of-sample rolling horizon weighted errors. *Expert Systems with Applications*, **38**, 14778–14785.

R Core Team (2013). *R: A Language and Environment for Statistical Computing*. R Foundation for Statistical Computing, Vienna, Austria. ISBN 3-900051-07-0.

Sayed, H. E., Gabbar, H. A. & Miyazaki, S. (2009). A hybrid statistical genetic-based demand forecasting expert system. *Expert Systems with Applications*, **36**, 11662–11670.

Tashman, L. J. (2000). Out-of-sample tests of forecasting accuracy: an analysis and review. *International Journal of Forecasting*, **16**(4), 437–450.

Thomakos, D. & Nikolopoulos, K. (2014). Fathoming the theta method for a unit root process. *IMA Journal of Management Mathematics*, **25**, 105–124.

Winters, P. R. (1960). Forecasting sales by exponentially weighted moving averages. *Management Science*, **6**, 324–342.